\documentclass{eas}
\usepackage{graphicx}
\begin{document}

\title{Binary Cepheids from optical interferometry} 
\author{A.~Gallenne}\address{Universidad de Concepci\'on, Departamento de Astronom\'ia, Casilla 160-C, Concepci\'on, Chile ; \email{agallenne@astro-udec.cl}}
\author{P.~Kervella}\address{LESIA, Observatoire de Paris, CNRS UMR 8109, UPMC, Universit\'e Paris Diderot, 5 Place Jules Janssen, F-92195 Meudon, France}
\author{A.~M\'erand}\address{European Southern Observatory, Alonso de C\'ordova 3107, Casilla 19001, Santiago 19, Chile}
\author{J.~D.~Monnier}\address{Astronomy Department, University of Michigan,1034 Dennison Bldg, Ann Arbor, MI 48109-1090, USA}
\author{J.~Breitfelder}\sameaddress{2,3}
\author{G.~Pietrzynski}\address{Warsaw University Observatory, Al. Ujazdowskie 4, 00-478 Warsaw, Poland}\sameaddress{,1}
\author{W.~Gieren}\sameaddress{1}

\begin{abstract}

Classical Cepheid stars have been considered since more than a century as reliable tools to estimate distances in the universe thanks to their Period-Luminosity (P-L) relationship. Moreover, they are also powerful astrophysical laboratories, providing fundamental clues for studying the pulsation and evolution of intermediate-mass stars. When in binary systems, we can investigate the age and evolution of the Cepheid, estimate the mass and distance, and constrain theoretical models. However, most of the companions are located too close to the Cepheid ($\sim 1$-40 mas) to be spatially resolved with a 10-meter class telescope. The only way to spatially resolve such systems is to use long-baseline interferometry. Recently, we have started a unique and long-term interferometric program that aims at detecting and characterizing physical parameters of the Cepheid companions, with as main objectives the determination of accurate masses and geometric distances.

\end{abstract}
\maketitle

\section{Introduction}

Cepheids in binary systems are important tools to measure fundamental parameters. The dynamical masses can be estimated (Pietrzy\'nski et al. \cite{Pietrzynski_2011_12_0}, \cite{Pietrzynski_2010_11_0} ; Evans et al. \cite{Evans_2008_09_0}), providing new constraints on evolution and pulsation theory (see e.g. Prada Moroni et al. \cite{Prada-Moroni_2012_04_0}). This gives new insight on the Cepheid mass, and has the potential to settle the discrepancy between pulsation and evolution models. For many years, stellar evolutionary models have predicted Cepheid masses larger than those derived from pulsation models (see e.g. Neilson et al. \cite{Neilson_2011_05_0}).

Binary systems are also an excellent tool to determine independent distance measurements of Cepheids, needed to calibrate the P--L relation. For the Galactic Cepheids, the combination of radial velocity (RV) and astrometric measurements can provide orbital parallaxes and masses. However, the sensitivity of the Hipparcos telescope was insufficient to obtain astrometric orbital solutions, which prevented us from deriving any accurate distance measurements. In addition, all measured binary Cepheids are single-line spectroscopic binaries (for now), making the task even more difficult.

A number of companions of Galactic Cepheids were detected from radial velocity measurements and the variability of the $\gamma$-velocity of the Cepheids (see e.g. Moore \cite{Moore_1929_02_0} ; Herbig \& Moore \cite{Herbig_1952_09_0} ; Abt \cite{Abt_1959_11_0} ; Szabados \cite{Szabados_1989_01_0}, \cite{Szabados_1991_01_0}, and references therein). An ultraviolet survey was also carried out with the  International Ultraviolet Exporer (IUE), where companions were detected from low- and high-resolution spectra (Bohm-Vitense \& Proffitt \cite{Bohm-Vitense_1985_09_0} ; Evans \cite{Evans_1992_01_0}), providing us with a range of spectral types. From an evolutionary time-scale point of view, most of the companions should be stars close to the main sequence, and because of the Cepheid's brightness, only bright (and hence massive) companions can be detected from photometric or spectroscopic surveys. Fainter (and hence less massive) companions have a small effect on the Cepheid's astrometry, and might be detected from high-precision radial velocity measurements. However, because of non-symmetric lines from the Cepheid's atmosphere, a precision of the order of 1\,m~s$^{-1}$ is not possible.

Most of the secondary stars are located too close to the Cepheid ($\sim 1$-40\,mas) to be observed with a 10-m class telescopes. The only actual way to spatially resolve such systems is to use long-baseline interferometry. Therefore, we started in 2012 a new and unique long-term interferometric program that aims at measuring astrometric positions of the brightest companions. The first goal is to determine the angular separation and the apparent brightness ratio from the interferometric visibility and closure phase measurements. Our long-term objective, which needs a good sampling of the orbital period, is to determine the full set of orbital elements (including $a, i$ and $\Omega$, unknown from single-line radial velocity measurements), absolute masses and geometric distances. To reach that goal, we have also started a spectroscopic observing campaign (CORALIE, SOPHIE, HST/STIS), which aims at completing archival spectroscopic data in order to have a large time coverage of the orbital periods which usually span years. But the primary goal of these spectroscopic observations is to extract the radial velocity of the companions from high signal-to-noise ratio spectra, which we will combine to our interferometric measurements.


\section{The close high-contrast companions}

Due to the brightness of the Cepheids, it is difficult to detect the companions from optical/infrared photometry and spectra. The hottest ones can be detected from UV spectra, as reported by Evans \cite{Evans_1992_01_0}, who measured spectral type typically ranging from B to A type main-sequence stars. In the optical (and longer wavelengths) the flux of the Cepheid dominates, making difficult a direct image and radial velocity measurements of the secondary components.

In addition, according to the radial-velocity orbital solutions and the parallax of the Cepheids, the companions are located at only few milli-arcsecond (mas) from the primary, making their detection impossible with a single-dish telescope. So far, only five companions have been imaged (Polaris, $\eta$~Aql, $\delta$~Cep, V659~Cen and S~Nor, Evans et al. \cite{Remage-Evans_2013_07_0} ; Evans et al. \cite{Evans_2008_09_0}), at an angular projected separation greater than $0.6^{\prime\prime}$. All the other companions have an estimated projected semi-major axis less than 40\,mas.

We present in Table~\ref{table__companion_parameters} some informations about a few systems we have in our program. The listed minimum projected angular separations are derived using $a_1\,\sin i$ estimated from RV and the fact that $M_\mathrm{1} \geq M_\mathrm{2}$, therefore $a\,\sin i \geq 2\,a_1\,\sin i = a_\mathrm{min}\,\sin i$. Such angular separations are only reachable using interferometry. In addition, interferometers usually work in the infrared domain, therefore we also need an instrument with a high sensitivity to reach contrasts of a few percent ($f_2/f_1 < 5$\,\% in $H$).

\begin{table}[!h]
\centering
\caption{Physical properties of some Cepheid companions.}
\begin{tabular}{ccccccc} 
\hline
\hline
Cepheid ($P_\mathrm{puls}$)	&	Sp.~Type	&	$P_\mathrm{orb}$	&	$\Delta V$	&	$\Delta H$	&	$a_\mathrm{min}\,\sin i$	&	$d$ 	\\
				(day)							&						&	(yr)							&		(mag)		&	(mag)			&	(mas)					&	(pc)		\\
\hline
V1334~Cyg (3.33)					&	B7.0V				&	5.3								&	2.3					&	3.8				&	7.2						&	683		\\
AW~Per (6.46)							&	B8.3V				&	40.0							&	2.2					&	5.1				&	32.4				&	853		\\
U~Aql (7.02)							&	  B9.8V					&	5.1							&	3.4					&	5.8				&		4.0					&	668		\\
AX~Cir (5.27)							&	B6.0V					&	17.9						&	   1.9				&	4.3				&		24.2				&	500		 \\
S~Mus (9.66)							&	B3.5V					&	1.4							&	2.0					&	4.4				&	1.6						&	847		\\
V636~Sco (6.80)						&	B9.5V					&	3.7							&	3.4					&	5.6				&	3.8						&	823		\\
V350~Sgr (5.15)						&	B9.0V					&	4.1							&	2.7					&	5.1				&	3.0						&	920		\\
\hline
&	\\
\multicolumn{7}{p{.9\textwidth}}{Notes. Spectral types taken from Evans (\cite{Evans_1995_05_0}). Estimated distances from P--L relations (Storm et al. \cite{Storm_2011_10_0} ; Bono et al. \cite{Bono_2002_07_0}). Orbital elements taken from the database of Szabados (\cite{Szabados_2003_03_0}). Magnitude differences estimated from the spectral types using Table~15.7 of Cox (\cite{Cox_2000__0}) and intrinsic colors of Ducati et al. (\cite{Ducati_2001_09_0}).}
\end{tabular}
\label{table__companion_parameters}
\end{table}

\section{Basics of optical interferometry}

Interferometers combine the light coming from at least two telescopes to obtain interference fringes. The provided angular resolution no longer depends on the diameter of the telescope, but on the baseline, $B$,  between telescopes. For instance, we can reach a resolution down to 0.2\,mas in $V$ with $B = 330$\,m.

In practice, we measure the amplitude and phase of the interference fringes to form the complex visibility, $\tilde{V}_\mathrm{ij}(u,v)$, with $u = B_\mathrm{x}/\lambda$ and $v = B_\mathrm{y}/\lambda$. This quantity is directly proportional to the Fourier transform of the object brightness distribution (Van Cittert-Zernike Theorem). This is the basis of the image reconstruction method: 
\begin{displaymath}
\tilde{V}_\mathrm{ij}(u,v) = |V_\mathrm{ij}| e^\mathrm{i\varphi} \propto \hat{I}(\alpha,\delta),
\end{displaymath}
where the tilde and hat symbol denote, respectively, a complex quantity and a Fourier transform, while the indexes stand for telescope $i$ and $j$. However, one interferometric measurement (i.e. for a given baseline) provides only one Fourier component of the intensity distribution. We therefore need to collect sufficient Fourier components to improve our knowledge of the source brightness distribution. That is why to increase the $(u,v)$ plane coverage and minimize the observing time, we use several telescopes (with different baseline orientations) and spectral dispersion.

To remove/reduce various sources of noise, the "real" observables in interferometry are the squared visibilities, $V^2$, and the closure phases, $\phi$. The squared visibility is related to the fringe amplitude (fringe contrast), $V^2 = |V_\mathrm{ij}|^2$, and is sensitive to the angular size of the source. The closure phase, $\phi_{ijk}$, more sensitive to asymmetric structures, is formed through the triple product of the complex visibility around a close triangle, i.e. we form the bispectrum:
\begin{displaymath}
\tilde{B}_\mathrm{ijk} = \tilde{V}_\mathrm{ij} \tilde{V}_\mathrm{jk} \tilde{V}_\mathrm{ki} = |V_\mathrm{ij}| |V_\mathrm{jk}| |V_\mathrm{ki}| e^\mathrm{\varphi_\mathrm{ij} + \varphi_\mathrm{jk} + \varphi_\mathrm{ki}} = |V_\mathrm{ijk}| e^\mathrm{i\phi_{ijk}},
\end{displaymath}
where the quantity $|V_\mathrm{ijk}|$ is called the amplitude of the triple product. As noticed, we need at least three telescopes to form the bispectrum.

When searching for high contrast companions, the main observable is the closure phase signal, because the variations in the squared visibility are almost undetectable. We therefore need at least a three-telescope recombiner allowing a high dynamic range. The only available instruments with such capabilities are CHARA/MIRC (Michigan InfraRed Combiner) and VLTI/PIONIER (Precision Integrated-Optics Near-infrared Imaging ExpeRiment).

\section{The MIRC and PIONIER recombiners}

MIRC (Monnier et al. \cite{Monnier_2004_10_0}, \cite{Monnier_2010_07_0}) is installed at the CHARA array, located on Mount Wilson, California. The array consists of six 1\,m aperture telescopes with an Y-shaped configuration (two telescopes on each branch), oriented to the east, west and south, with baselines ranging from 34\,m to 331\,m. MIRC combines the light coming from all six telescopes in the $K$ or $H$ bands, with three spectral resolutions ($R = 42, 150$ and 400). The recombination of six telescopes gives simultaneously 15 fringe squared visibilities and 20 closure phase measurements for each spectral channel.

PIONIER (Le Bouquin et al. \cite{Le-Bouquin_2011_11_0}) is installed at the Very Large Telescope Interferometer (VLTI), located at Cerro Paranal, Chile (visitor instrument so far). The array consists of four fixed 8.2\,m Unit Telescopes (UTs), and four relocatable 1.8\,m Auxiliary Telescopes (ATs), providing baselines from 11\,m to 140\,m. PIONIER is a four-telescope combiner, and offers spectral dispersion in 1 (broad band), 3 or 7 spectral channels. The recombination provides simultaneously 6 visibilities and 4 closure phase signals per spectral channel.

\section{The short-period Cepheid V1334~Cyg}

\begin{figure}[!t]
\centering
 \resizebox{\hsize}{!}{\includegraphics{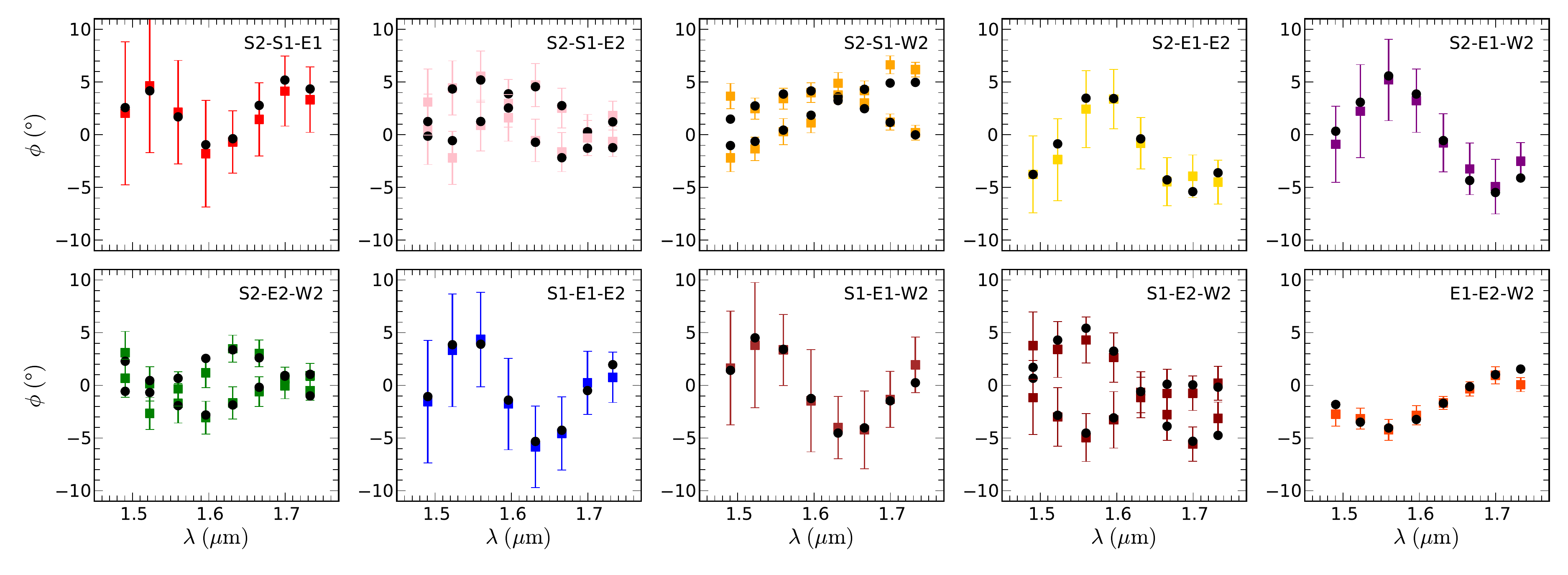}}
 \caption{Closure phase signal for the one epoch observation and using four telescopes. The color-coded squares are the data, while the black dots represent the fitted binary model (Gallenne et al. \cite{Gallenne_2013_04_0}). Note that in the absence of companion, the closure phase signal should be zero.}
   \label{image__cp}
\end{figure}

\begin{figure}[!t]
\centering
 \resizebox{\hsize}{!}{\includegraphics{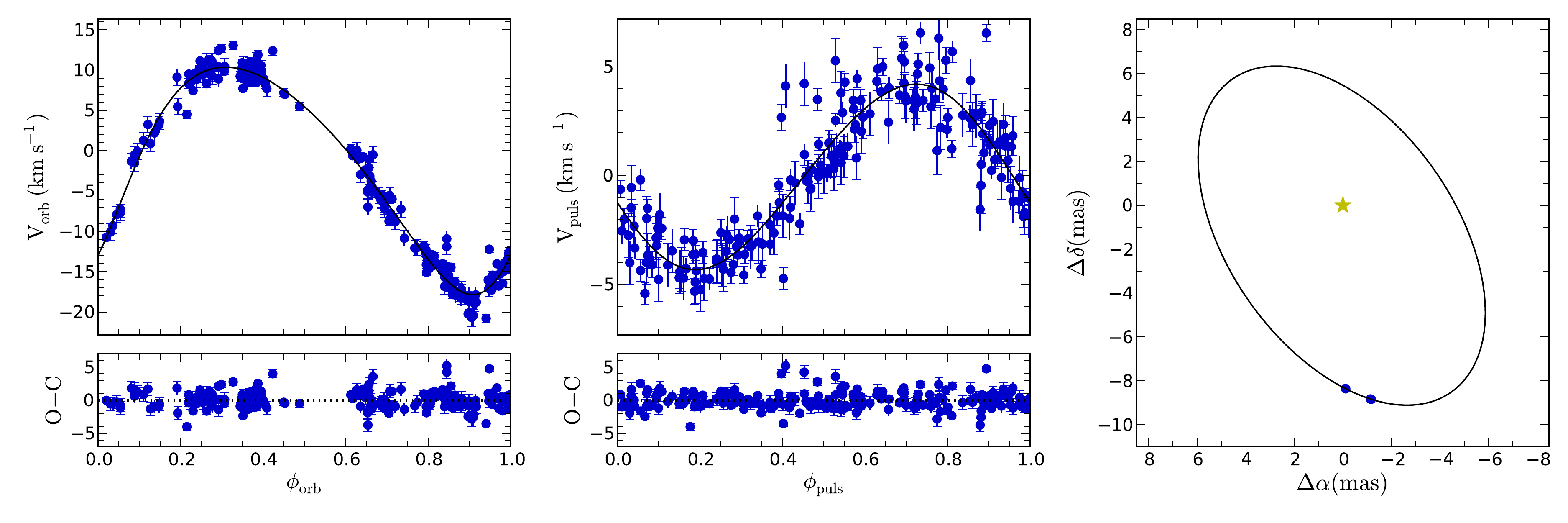}}
 \caption{\textit{Left}: fitted (solid line) and measured orbital velocity. \textit{Middle}: fitted (solid line) and measured pulsation velocity. \textit{Right}: orbit of V1334~Cyg Ab. The blue data points are the MIRC observations of 2012. The red point denotes our last measurements of 2013, and the green one the predicted value from the orbital elements derived in Gallenne et al. (\cite{Gallenne_2013_04_0}).}
   \label{image__orbit}
\end{figure}

Our binary Cepheid program started with MIRC observations, and the first dataset already provided very promising results. From the variations of the closure phase signal, showed in Fig.~\ref{image__cp}, we detected the companion orbiting the short-period Cepheid V1334~Cyg. We measured its astrometric position and flux ratio with an accuracy better than 3\,\% (Gallenne et al. \cite{Gallenne_2013_04_0}). We then combined our astrometric data with available single-line radial velocity (RV) measurements to derive for the first time all orbital elements of the system. More particularly, we derived the orbital inclination, $i = 124.7 \pm 1.8^\circ$, the angular semi-major axis, $a = 8.54 \pm 0.51$\,mas, and the longitude of the ascending node,  $\Omega = 206.3 \pm 9.4^\circ$ which were previously unknown. The final orbital fit is shown in Fig.~\ref{image__orbit}.

From our derived flux ratio, we also estimated the apparent magnitude of the companion, $m_\mathrm{H} = 8.47 \pm 0.15$\,mag. However, we were not able to estimate the distance and mass of the Cepheid as no  RV measurements of the companion exist yet (it is a single-line spectroscopic binary). We therefore set lower limits on the masses, the spectral type of the companion, and the distance of the system.

In August 2013, we obtained a new point in the orbit in a very good agreement with the prediction. This is shown in Fig.~\ref{image__orbit} where the green dot denotes the predicted value, while the red one is our measurement. This demonstrates the accuracy of our preliminary orbital fit.

\section{Last new results: AW~Per and AX~Cir}

\begin{figure}[!t]
\centering
 \resizebox{\hsize}{!}{\includegraphics{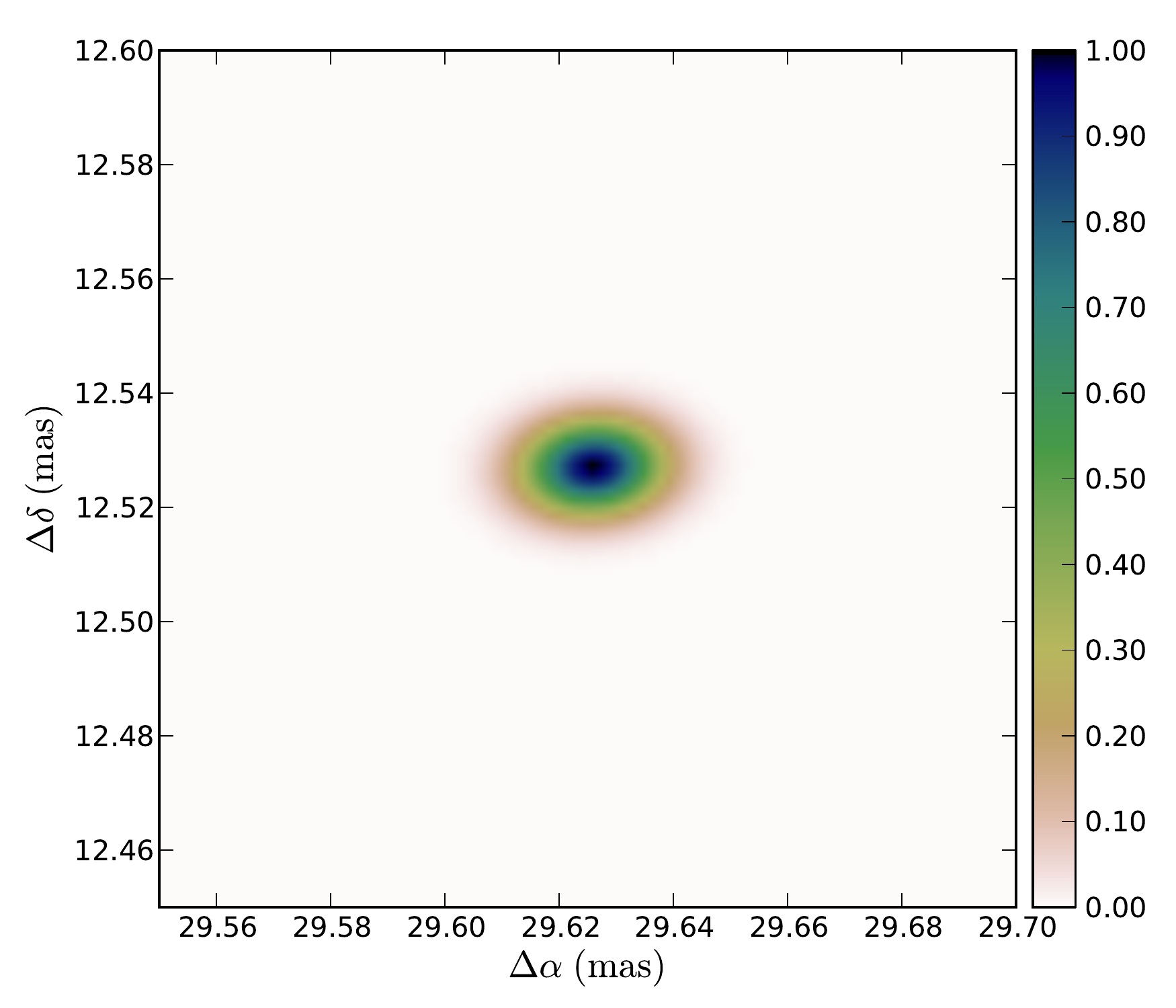}
 \includegraphics{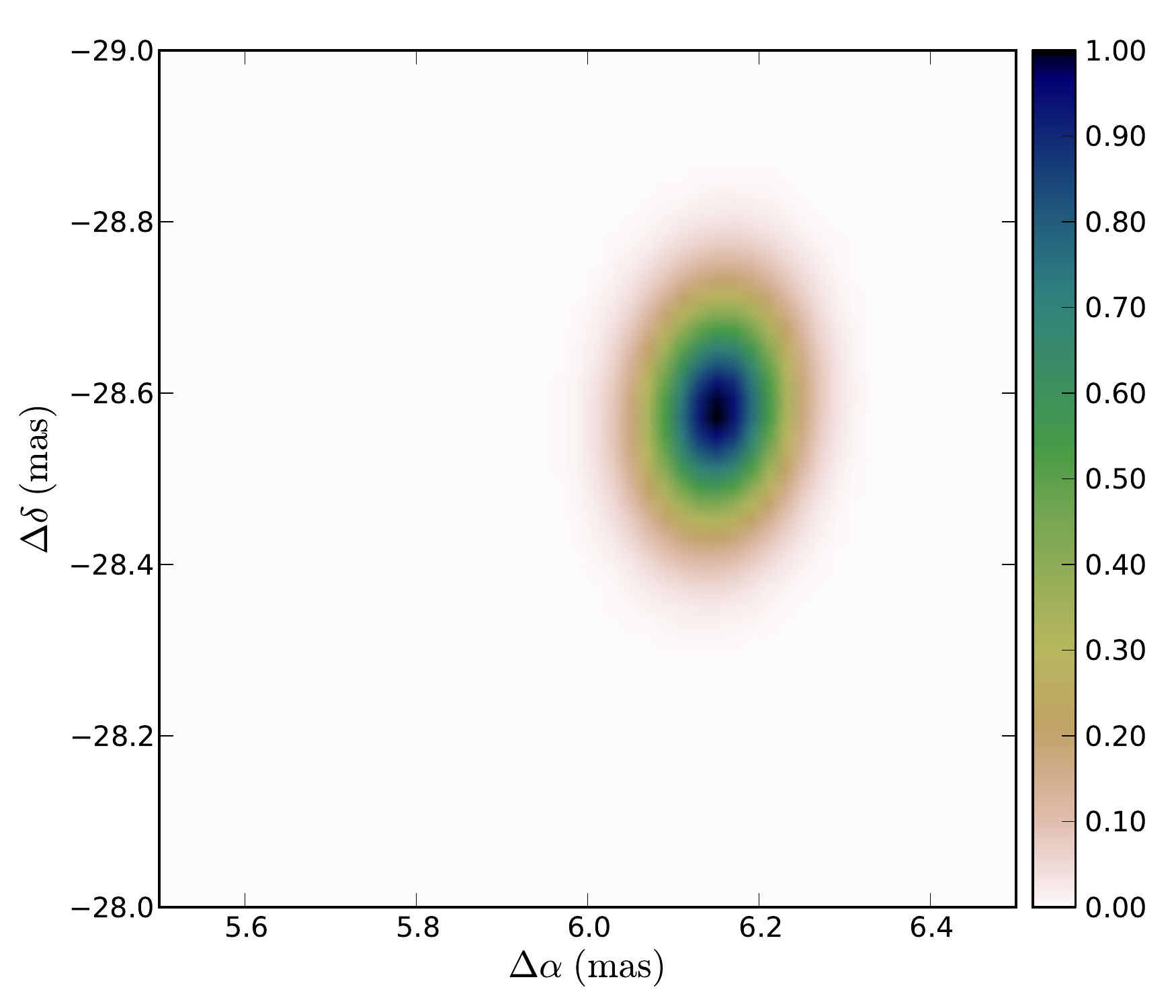}}
 \caption{Probability maps for the position of the companion of AX~Per (\textit{left}) and AX~Cir (\textit{right}), showing the maximum obtained at each point of the search region.}
   \label{image__chi2maps}
\end{figure}

Recently, we also reduced the data for the Cepheids AW~Per and AX~Cir, respectively observed with MIRC and PIONIER. The companion is detected for both stars at a projected angular separation $\sim 30$\,mas. We present the probability maps in Fig.~\ref{image__chi2maps}. For AX~Cir, the companion is detected at a projected angular separation $\rho = 29.2$\,mas and a position angle $PA = 167.6^\circ$, with a flux ratio between the companion and the Cepheid $f = 0.90$\,\% (Gallenne et al. 2013, submitted to A\&A).

The most probable location of the companion of AW~Per is at $\rho = 32$\,mas and $PA = 67^\circ$. We also measure a magnitude difference $\Delta H = 4.57$\,mag.

\begin{figure}[!t]
\centering
 \resizebox{\hsize}{!}{\includegraphics{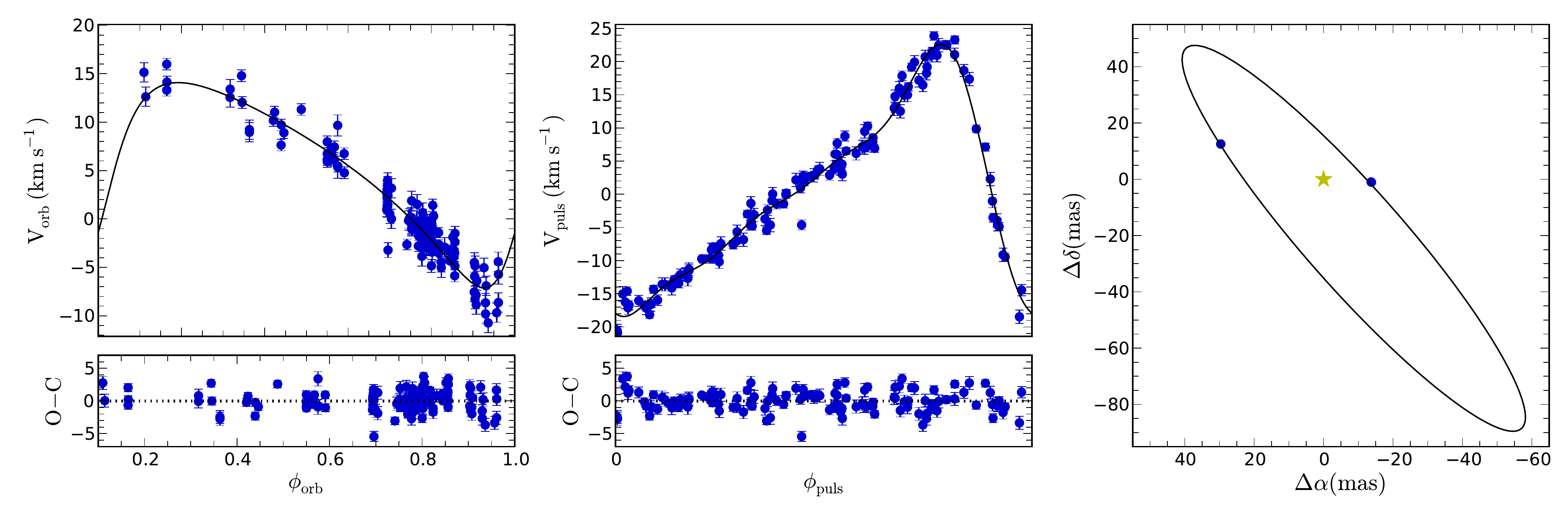}}
 \caption{\textit{Left}: fitted (solid line) and measured orbital velocity. \textit{Middle}: fitted (solid line) and measured pulsation velocity. \textit{Right}: orbit of AW~Per Ab. The blue data points are the MIRC observations of 2012 and the HST measurement of 2001.}
   \label{image__orbit_awper}
\end{figure}

We can even go further for AW~Per by applying the same analysis than V1334~Cyg as an additional astrometric measurement is available in the literature (Massa \& Evans \cite{Massa_2008_01_0}). We therefore collected all RV measurements (single-line) that we combined with the astrometric position. The result is plotted in Fig.~\ref{image__orbit_awper}. Our fitted orbital parameters are very consistent with the spectroscopic orbital elements derived by Evans et al. (\cite{Evans_2000_07_0}). Additionally, we estimate a semi-major axis $a \sim 88$\,mas, an inclination angle $i \sim 80^\circ$, and the longitude of the ascending node $\Omega = 100^\circ$ (Gallenne et al. 2014, in prep.).

\section{Conclusion}

Thanks to the high angular resolution provided by interferometry, we are able to detect the close companions of Cepheids. Our work, using multi-telescope recombination, provided novel physical parameters of the Cepheid companions. These innovative results also show the capabilities of long-baseline interferometry to study close and high-contrast binary systems in general.

The astrometric measurements provided by interferometry can then be efficiently combined with radial velocity data to provide new constraints on the system properties. Further interferometric observations will be obtained in the future to cover the orbit, and derive the orbital elements with a unique accuracy for our sample of binary Cepheids.

Finally, our interferometric program is complementary to an ongoing spectroscopic program to detect the orbital radial velocity variations induced by the companions. This will provide an orbital parallax and model-free masses for Galactic Cepheids.

\begin{acknowledgements}
AG acknowledges support from FONDECYT grant 3130361. JDM acknowledges funding from the NSF grants AST-0707927 and AST-0807577. WG and GP gratefully acknowledge financial support for this work from the BASAL Centro de Astrof\'isica y Tecnolog\'ias Afines (CATA) PFB-06/2007. Support from the Polish National Science Centre grant MAESTRO DEC-2012/06/A/ST9/00269 and the Polish Ministry of Science grant Ideas Plus (awarded to GP) is also acknowledge. We acknowledge financial support from the “Programme National de Physique Stellaire” (PNPS) of CNRS/INSU, France. This research received the support of PHASE, the high angular resolution partnership between ONERA, Observatoire de Paris, CNRS, and University Denis Diderot Paris 7. The research leading to these results received funding from the European Research Council under the European Community's Seventh Framework Programme (FP7/2007--2013)/ERC grant agreement n$^\circ$227224 (PROSPERITY).
\end{acknowledgements}



\begin{thebibliography}{99}

\bibitem[1959]{Abt_1959_11_0} Abt, H. A. 1959, ApJ, 130, 769
\bibitem[1985]{Bohm-Vitense_1985_09_0} Bohm-Vitense, E. \& Proffitt, C. 1985, ApJ, 296, 175
\bibitem[2002]{Bono_2002_07_0} Bono, G., Groenewegen, M. A. T., Marconi, M., \& Caputo, F. 2002, ApJ, 574, L33
\bibitem[2000]{Cox_2000__0} Cox, A. N. 2000, Allen's astrophysical quantities
\bibitem[2001]{Ducati_2001_09_0} Ducati, J. R., Bevilacqua, C. M., Rembold, S. B., \& Ribeiro, D. 2001, ApJ, 558, 309
\bibitem[1992]{Evans_1992_01_0} Evans, N. R. 1992, ApJ, 384, 220
\bibitem[1995]{Evans_1995_05_0} Evans, N. R. 1995, ApJ, 445, 393
\bibitem[2000]{Evans_2000_07_0} Evans, N. R., Vinko, J., \& Wahlgren, G. M. 2000, AJ, 120, 407
\bibitem[2008]{Evans_2008_09_0} Evans, N. R., Schaefer, G. H., Bond, H. E., et al. 2008, AJ, 136, 1137
\bibitem[2013]{Gallenne_2013_04_0} Gallenne, A., Monnier, J. D., M\'erand, A., et al. 2013, A\&A, 552, A21
\bibitem[2011]{Le-Bouquin_2011_11_0} Le Bouquin, J.-B., Berger, J.-P., Lazareff, B., et al. 2011, A\&A, 535, A67
\bibitem[2013]{Remage-Evans_2013_07_0} Evans, N. R, Bond, H. E., Schaefer, G. H., et al. 2013, in press [ArXiv e-prints: 1307.7123]
\bibitem[1959]{Herbig_1952_09_0} Herbig, G. H. \& Moore, J. H. 1952, ApJ, 116, 348
\bibitem[2010]{Pietrzynski_2010_11_0} Pietrzy\'nski, G., Thompson, I. B., Gieren, W., et al. 2010, Nature, 468, 542
\bibitem[2011]{Pietrzynski_2011_12_0} Pietrzy\'nski, G., Thompson, I. B., Graczyk, D., et al. 2011, ApJ, 742, L20
\bibitem[2012]{Prada-Moroni_2012_04_0} Prada Moroni, P. G., Gennaro, M., Bono, G., et al. 2012, ApJ, 749, 108
\bibitem[2008]{Massa_2008_01_0} Massa, D. \& Evans, N. R. 2008, MNRAS, 383, 139
\bibitem[2004]{Monnier_2004_10_0} Monnier, J. D., Berger, J.-P., Millan-Gabet, R., \& ten Brummelaar, T. A. 2004, in SPIE Conference Series, ed. W. A. Traub, Vol. 5491, 1370
\bibitem[2010]{Monnier_2010_07_0} Monnier, J. D., Anderson, M., Baron, F., et al. 2010, in SPIE Conference Series, ed. W. C. Danchi, F. Delplancke, \& J. K. Rajagopal, Vol. 7734, 12
\bibitem[1929]{Moore_1929_02_0} Moore, J. H. 1929, PASP, 41, 56
\bibitem[2011]{Neilson_2011_05_0} Neilson, H. R., Cantiello, M., \& Langer, N. 2011, A\&A, 529, L9
\bibitem[2011]{Storm_2011_10_0} Storm, J., Gieren, W., Fouqu\'e, P., et al. 2011, A\&A, 534, A94
\bibitem[1989]{Szabados_1989_01_0} Szabados, L. 1989, Commmunications of the Konkoly Observatory Hungary, 94, 1
\bibitem[1991]{Szabados_1991_01_0} Szabados, L. 1991, Commmunications of the Konkoly Observatory Hungary, 96, 123
\bibitem[2003]{Szabados_2003_03_0} Szabados, L. 2003, Information Bulletin on Variable Stars, 5394, 1

\end{thebibliography}
\end{document}